\documentclass[letterpaper,10pt]{article} 
\usepackage{osameet3} %% use version 3 for proper copyright statement

\newcommand\authormark[1]{\textsuperscript{#1}}

\usepackage{amsmath,amssymb}
\usepackage[colorlinks=true,bookmarks=false,citecolor=blue,urlcolor=blue]{hyperref} %pdflatex
\usepackage{pgfplots}
\usepackage{filecontents}
\usepackage{amsmath,amssymb}
\usepackage{subcaption}
\usepackage{tikz}
\usetikzlibrary{quantikz}
\tikzset{block/.style={draw, thick, text width=1.5cm, minimum height=0.5cm, align=center}}

\begin{document}

\title{Case study on quantum convolutional neural network scalability}

% \author{Author name(s)}
% \address{Author affiliation and full address}
% \email{e-mail address}
%%Uncomment the following line to override copyright year from the default current year.
%\copyrightyear{2022}

\author{Marina O. Lisnichenko\authormark{1},  Stanislav I. Protasov\authormark{1}}%Author Two,\authormark{2,*} and Author Three\authormark{2,3}}

\address{\authormark{1} Machine Learning and Knowledge representation laboratory, Innopolis University, Innopolis, Russia}

% \authormark{2}Publications Department, Optica Publishing Group, 2010 Massachusetts Avenue NW, Washington, DC 20036\\
% \authormark{3}Currently with the Department of Electronic Journals, Optica Publishing Group, 2010 Massachusetts Avenue NW, Washington, DC 20036}
\email{m.lisnichenko@innopolis.university, s.protasov@innopolis.ru} %% email address is required

%Real world tasks utilize convolutional neural networks as a solution. Binary classification is one of the application examples. 

% The neural network processes an input image by mathematical operations such as convolution and pooling. After processing], the neural network ideally returns a single value -- class prediction. Model trains with a respect to an error between prediction and ground-truth. This training pipeline and its modifications are well-known nowadays for the classical computers. At the same time, quantum computers, hence, quantum machine learning take place in researches. But are quantum neural networks promising in the current stage? 

%It means that currently it is not possible to scale quantum convolutional neural networks for the projects which imply the use of increased input data, that are usual for real-life tasks.

\begin{abstract}
One of the crucial tasks in computer science is the processing time reduction of various data types, i.e., images, which is important for different fields - from medicine and logistics to virtual shopping. Compared to classical computers, quantum computers are capable of parallel data processing, which reduces the data processing time. This quality of quantum computers inspired intensive research of the potential of quantum technologies applicability to real-life tasks. Some progress has already revealed on a smaller volumes of the input data. In this research effort, I aimed to increase the amount of input data (I used images from $2\times2$ to $8\times8$), while reducing the processing time, by way of skipping intermediate measurement steps. The hypothesis was that, for increased input data, the omitting of intermediate measurement steps after each quantum convolution layer will improve output metric results and accelerate data processing. To test the hypothesis, I performed experiments to chose the best activation function and its derivative in each network. The hypothesis was partly confirmed in terms of output mean squared error (MSE) - it dropped from 0.25 in the result of classical convolutional neural network (CNN) training to 0.23 in the result of quantum convolutional neural network (QCNN)  training. In terms of the training time, however, which was 1.5 minutes for CNN and 4 hours 37 minutes in the least lengthy training iteration, the hypothesis was rejected. 
\end{abstract}

\section{Introduction}
Classical (i.e., implemented on a personal computer) convolutional neural network (CNN) is a universal approximation technique used in research and practical tasks~\cite{ameisen2020building}. For instance, classical convolutional neural networks are used in the binary image classification task \cite{alparslan2021towards}. 

Quantum programming branch is rapidly developing \cite{national2019quantum}. Special quantum operators named \textit{gates} allow to write programs. Quantum computations posses unique properties \cite{nielsen2002quantum}. Potentially, these properties allow quantum computers to solve tasks faster than classical computers \cite{williams1998explorations, biamonte2017quantum}.

Hence, the aim of this study was to assemble quantum convolutional neural network (QCNN) for the binary classification task.

As a foundation, current work compares already existing QCNN implementations. The wide-spread strategy is to mimic the classical CNN. For instance, multiple solutions for convolution and pooling layers exist. I modify the observed models to achieve the most prominent training pipeline. The section below observes the reviewed solutions.

The rest of this paper takes a form of four sections. Section \ref{sec:problem} provides information about case-study dataset. 

Section \ref{sec:solution}) describes the solution.
Section \ref{sec:quantum_gates} describes quantum gates used in the current work.
Section \ref{sec:data_ecoding} delivers image encoding information.
Section~\ref{sec:conv_and_pool} represents a QCNN example circuit.
Section \ref{sec:measurement_and_error} describes an activation function used in QCNN.
Section \ref{sec:back_propagation} provides detailed information on back-propagation step.
Part \ref{sec:parameters} provides the QCNN parameters information. These parameters are common for all the experiments except if they are not indicated. Section \ref{sec:results} brings the training process with mentioned parameters results. Conclusion \ref{sec:discuss_and_conclude} includes results discussion and possible further improvements.

\section{Methods}
\label{sec:methods}
The problem solution contains the following steps:

\begin{enumerate}
    \item data encoding;
    \item forward pass;
    \item error calculation;
    \item backward pass.
\end{enumerate}

Data encoding step assumes transformation of classical data into quantum. In my study the data are classical images. In the previous work \cite{lisnichenkoquantum} we discussed possible image representation techniques for quantum computer. The Qubit Lattice approach \cite{venegas2003storing} encodes image in the closest to the classical representation manner. The related work observation showed that authors used classical-like approaches to construct quantum convolutional neural network (QCNN). Therefore, I consider the Qubit Lattice as a relevant choice to encode image.

Forward pass step is responsible for encoded image processing. Any artificial (quantum or classical) neural network processes data with mathematical operations. In case of image processing, these operations are named \textit{convolution} and \textit{pooling}. Part \ref{sec:rel_work} describes already existing quantum convolution and pooling implementations. The result of the convolution-pooling processing in the current work is a single qubit. I hypothesise that this wire contains information about input image label prediction.

The error calculation step provides information about prediction correctness. In case of quantum computations, output qubit is measured primarily. After that, another function called activation handles the measured value. If the handled value exceeds the classification threshold value, the prediction label equals $1$. In the opposite case, the label equals $0$. Label prediction is compared with ground-truth label to estimate the error.

Backward step trains the model based on the prediction error. The step is the most important part of this research. In this study the step comprises three items.

Firstly, I explore three derivative calculation algorithms and determine which one performs better. As mentioned earlier, CNN consists of convolution and pooling layers. These layers allow to subtract features that are important for prediction. After processing, these features pass through the activation function. This function introduces non-linearity that computes the importance of extracted features. Taking this fact into account I introduce three back-propagation algorithms. Three derivative calculation approaches correspond to these algorithms.

Secondly, I analyse the effort of intermediate measurements on the output prediction. The back-propagation algorithm is known for classical computers \cite{bouvrie2006notes}. Having the current parameters values and calculated output error, the algorithm updates the weights of the layers. The opposite situation is for the quantum hyper-parameterized layers. Strictly speaking, the network has quantum nature until it is measured. Measurement collapses quantum state into classical \cite{leggett2005quantum}. To determine a classical value, circuit executes multiple times. Multiple measurements provide layer outcome based on classical values statistics. This execution-measurement procedure is costly in case of a single layer. For this reason, my experiment aims to check the network results based on training with and without intermediate measurements.

Finally, I observe layer-wise and simultaneous parameter updating processes. Based on the previous observations, I build two multi-layer QCNN. The first neural network updates its weights simultaneously with a single gradient for all the layers. While the second neural network trains with layer-wise approach, where gradient is calculated separately for every layer.

\section{Related Work}
\label{sec:rel_work}
Mostly researchers either mimic the classical convolution layers by quantum mathematics or introduce them in pair with classical layers. For example, authors \cite{henderson2020quanvolutional} described influence of quantum convolution layers on classical neural networks. They performed a quantum kernel as a quantum random circuit with each wire measured separately. The activation extracts a single value from the measured wires. This value is a convolution layer outcome.

Authors used three training models:

\begin{enumerate}
    \item classical with convolution, pooling and fully connected layers;
    \item classical model with quantum convolution first layer replacement;
    \item classical model with "purely classical random non-linear transformation" first layer replacement.
\end{enumerate}

Authors showed that quantum layer replacement refines the test accuracy and training loss. Convolution replacement by random classical kernel leads to the same improvements. The paper graphically explains the relation between replaced classical to quantum layers number and output metrics. The greater number of quantum layers the higher results are. However, paper does not describe back-propagation process and training parameters to confirm the research empirically.

Iris Cong \textit{et al.} \cite{cong2019quantum} maintained a quantum-classical (hybrid) neural network. The multi-scale entanglement re-normalization ansatz (MERA) \cite{awschalom2013quantum} forms a network model's base. QCNN helped to solve the many-body problem task.

\begin{figure}[!ht]
    \centering
        \scalebox{0.75}
        {
            \begin{quantikz}[scale=0.1]
                \lstick{(0, 0)} & \gate{RX(a_{00})} & \gate{RZ} & \gate{RX} & \gate{RZ} & \gate{RX} & \meter{} \\
                \lstick{(0, 1)} & \gate{RX(a_{01})} & \ctrl{-1} & \ctrl{-1} & \rstick{}\\
                \lstick{(1, 0)} & \gate{RX(a_{10})} & \gate{RZ} & \gate{RX} & \ctrl{-2} & \ctrl{-2} & \rstick{}\\
                \lstick{(1, 1)} & \gate{RX(a_{11})} & \ctrl{-1} & \ctrl{-1} & \rstick{}\\
            \end{quantikz}
        }
    \caption{Convolution layer\cite{oh2020tutorial}}
    \label{fig:conv}
\end{figure}
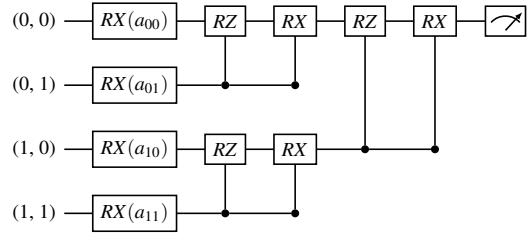

Researches \cite{oh2020tutorial} applied the previous approach with modified convolution layer for image processing. Paper explains quantum convolution and pooling operations schematically as in the figure~\ref{fig:conv}.

Pooling layer shrinks a qubit number applying quantum error correction. Quantum control gate with removed measurement simplifies pooling operation. I exploit mentioned convolution and pooling operations with modifications.
As a result, authors claimed that test loss is 0.18 for quantum and 0.21 for classical models. The accuracy increased from 0.950 to 0.976. Additionally, paper provides such network training information as filter size, epoch number and image encoding. However, back-propagation processing description does not take place.

\begin{figure}[!ht]
    \centering
        \scalebox{0.75}
        {
            \begin{quantikz}
                \lstick{$\ket{0}$} & \gate{R_y(\arctan(x_1))} & \gate{R_z(\arctan(x_1^2))} & \ctrl{1} \gategroup[wires=4,steps=4,style={dashed, rounded corners}]{{}} & \qw & \qw & \gate{R(\alpha_1, \beta_1, \gamma_1))} & \meter{} & \qw \\
                \lstick{$\ket{0}$} & \gate{R_y(\arctan(x_2))} & \gate{R_z(\arctan(x_2^2))} & \targ{0} & \ctrl{1} & \qw & \gate{R(\alpha_2, \beta_2, \gamma_2))} & \qw & \qw \\
                \lstick{$\ket{0}$} & \gate{R_y(\arctan(x_3))} & \gate{R_z(\arctan(x_3^2))} & \qw & \targ{1} & \ctrl{1} & \gate{R(\alpha_3, \beta_3, \gamma_3))} & \qw & \qw \\
                \lstick{$\ket{0}$} & \gate{R_y(\arctan(x_4))} & \gate{R_z(\arctan(x_4^2))} & \qw & \qw & \targ{2} & \gate{R(\alpha_4, \beta_4, \gamma_4))} & \qw & \qw
            \end{quantikz}
        }
    \caption{Convolution layer \cite{chen2022quantum}. The set of first two gates for each wire stands for classical-to-quantum data representation; sub-circuit in the dashed line is a convolution itself. Tree parameters $\alpha, \beta, \gamma$ parameterize the convolution layer.}
    \label{fig:conv_chen}
\end{figure}
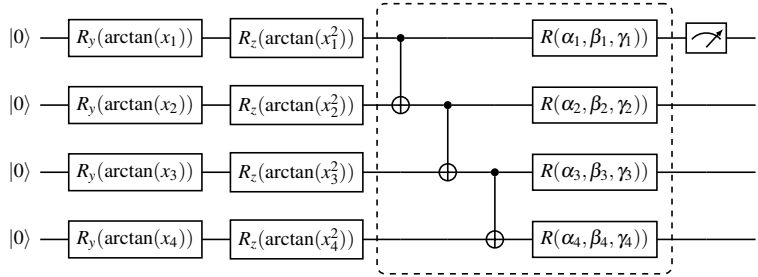

Chen \textit{et al.} \cite{chen2022quantum} constructed convolution kernel shown in the figure~\ref{fig:conv_chen}.

In the binary classification task authors achieved $97.5\%$ accuracy versus $82.5\%$ by classical CNN. Paper includes parameters and weights update technique using dynamic learning rate algorithm.

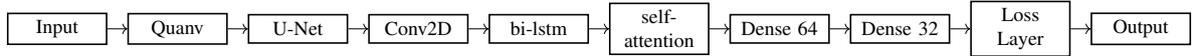
\begin{figure}[!ht]
    \centering
        \scalebox{0.75}
        {
            \begin{tikzpicture}[node distance=0.35cm,auto]
                \node[block] (a) {Input};
                \node[block,right=of a] (b) {Quanv};
                \node[block,right=of b] (c) {U-Net};
                \node[block,right=of c] (d) {Conv2D};
                \node[block,right=of d] (e) {bi-lstm};
                \node[block,right=of e] (f) {self-attention};
                
                \node[block,right=of f] (g) {Dense 64};
                \node[block,right=of g] (h) {Dense 32};
                \node[block,right=of h] (i) {Loss Layer};
                \node[block,right=of i] (j) {Output};
                \draw[->] (a) -- (b);
                \draw[->] (b) -- (c);
                \draw[->] (c) -- (d);
                \draw[->] (d) -- (e);
                \draw[->] (e) -- (f);
                \draw[->] (f) -- (g);
                \draw[->] (g) -- (h);
                \draw[->] (h) -- (i);
                \draw[->] (i) -- (j);
            \end{tikzpicture}
        }
    \caption{QCNN architecture, where Quanv stands for "Quantum convolution"; bi-LSTM stands for bi-directional long short-term memory and a self-attention encoder aims to get the best "spoken word recognition"}
    \label{fig:pipeline}
\end{figure}

Another report \cite{yang2021decentralizing} provides a pipeline (see figure~\ref{fig:pipeline}) for a speech recognition task. The approach combines quantum random layer and a U-Net network for the further processing.

\begin{figure}[!ht]
    \centering
        \scalebox{0.75}
        {
            \begin{quantikz}
                \lstick{$\ket{0}$} & \gate{R_y} & \gate{R_x} & \targ{3} & \gate{R_y} & \gate{R_x} & \meter{}\\
                \lstick{$\ket{0}$} & \gate{R_y} & \targ{2} & \qw & \qw & \qw & \meter{} \\
                \lstick{$\ket{0}$} & \gate{R_y} & \ctrl{-1} & \qw & \qw & \qw & \meter{} \\
                \lstick{$\ket{0}$} & \gate{R_y} & \gate{R_x} & \ctrl{-3} & \qw & \qw & \meter{}
            \end{quantikz}
        }
    \caption{Deployed quantum circuit \cite{yang2021decentralizing}}
    \label{fig:deployed_quantum_circuit}
\end{figure}
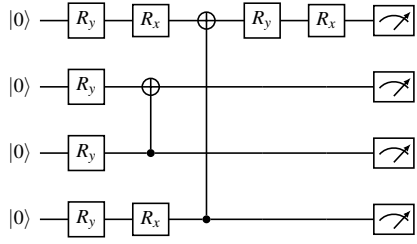

Authors provide random quantum circuit structure as in the figure~\ref{fig:deployed_quantum_circuit}. This structure follows the same logic as previous sources. However, instead of fully convoluted window of pixel-intensities, the result contains 4 values that are pushed to the further pipeline parts. %The classical analogy of the described operation can be $1\times1$ convolution \cite{iandola2016squeezenet}.
As authors claim, quantum layer improves the accuracy from 94.72 to 95.12.

\begin{figure}[!ht]
    \centering
        \scalebox{0.6}
        {
        \begin{subfigure}[b]{0.8\textwidth}
            \centering
            \begin{quantikz}
                \lstick{$\ket{a}$} & \qw                  & \gate[wires=2]{Conv} & \gate{Pool} & \qw      & \qw & \gate[wires=3, nwires=2]{Conv} \gategroup[wires=4,steps=1,style={inner sep=0.35cm, dashed, rounded corners, fill=white}, background]{{}} & \qw & \meter{} \\
                \lstick{$\ket{b}$} & \gate[wires=2]{Conv} &                      & \ctrl{-1}   & \meter{} & \cw & & \\
                \lstick{$\ket{c}$} & {}                   & \gate[wires=2]{Conv} & \gate{Pool} & \qw      & \qw & & \\
                \lstick{$\ket{d}$} & \qw                  &                      & \ctrl{-1}   & \meter{} & \cw & &
            \end{quantikz}
            \caption{Branching}
            \label{fig:branching_circ}
        \end{subfigure}
        }
        \hfill
        \scalebox{0.6}
        {
        \begin{subfigure}[b]{0.8\textwidth}
            \centering
            \begin{quantikz}
                \lstick{$i$} & \gate{U_3(\theta_0, \theta_1, \theta_2)} & \ctrl{1} & \gate{R_y(\theta_6)} & \targ{}   & \gate{R_y(\theta_8)} & \ctrl{1} & \gate{U_3(\theta_9, \theta_{10}, \theta_{11})}\\
                \lstick{$j$} & \gate{U_3(\theta_3, \theta_4, \theta_5)} & \targ{}  & \gate{R_z(\theta_7)} & \ctrl{-1} & \qw                  & \targ{}  & \gate{U_3(\theta_{12}, \theta_{13}, \theta_{14})}
            \end{quantikz}
            \caption{Convolution layer}
            \label{fig:_conv_box}
            \hfill
        \end{subfigure}
        }
    \caption{Branching circuit part~\ref{fig:branching_circ} with four input wires. Each \textit{Conv} box expands to the circuit~\ref{fig:_conv_box}}
    \label{fig:branching}
\end{figure}
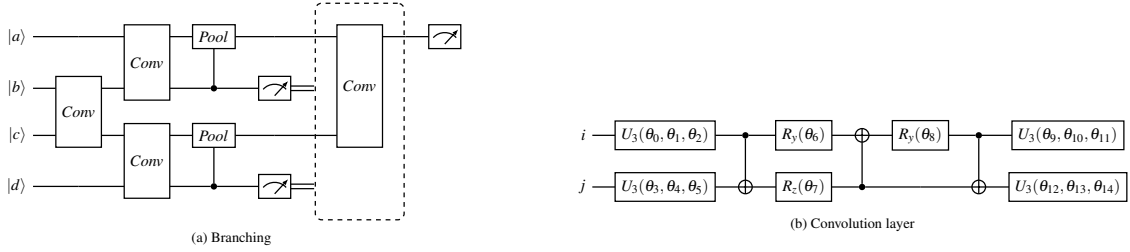

Another quantum convolution construction approach \cite{maccormack2022branching} explains how pooling measurements are used to affect on subsequent training process. Author named the choice of the next convolution layer "by the measurement outcomes of the previous pooling layer" as branching. Figure~\ref{fig:branching} represents the abstract training circuit.

The equation below expresses the $U_3$ gate.

\begin{equation}
    U_3(\theta_n, \theta_{n+1}, \theta_{n+2}) = 
    \begin{bmatrix}
    \cos{\frac{\theta_n}{2}} & -e^{i\theta_{n+2}}\sin{\frac{\theta_n}{2}} \\
    e^{i\theta_{n+1}}\sin\frac{\theta_n}{2} & e^{i\theta_{n+1}+i\theta_{n+2}}\cos{\frac{\theta_n}{2}}
    \end{bmatrix}
\end{equation}

Pooling operation consists of parameterized control rotation. As a result, the mean absolute error of the model is 0.9.

Kerenidis \textit{et al.} \cite{kerenidis2019quantum} developed a theoretical algorithm for deep QCNN. In this network each following layer receives measured wires from the previous layer as an input. Layer-wise measurements technique makes back-propagation algorithm similar to the classical implementation. Authors describe both forward and backward steps from the mathematical prospect.

Authors confirmed their pipeline numerically using MNIST dataset \cite{lecun1998gradient}. They claimed about training time speed-up and similar loss in comparison with classical convolutional neural network. Researchers provide such training information as data encoding procedure and parameters.

In this way, researchers mostly pay attention on how to simulate the classical CNN or make a hybrid from quantum and classical networks. The trick with layer-wise measurements allows to get access to the parameters on each layer. Likewise classical back-propagation algorithm updates weights, quantum rotation angles can be updated.

Wie \textit{et al.} provide an expectation value measurement method to update the parameters \cite{wei2022quantum}. The end-to-end circuit does not include intermediate measurements. The auxiliary qubits control the convolution layer choice. After convolution the pooling layer averages multiple wires into one. Authors suggest quantum fully connected layer idea as well. The equation below describes this idea as parameterized Hamiltonian.

\begin{equation}
    \mathcal{H} = h^0I + \sum_i{h^{i}\sigma_z^i} + \sum_{i, j}h^{i,j}\sigma_z^i\sigma_z^j, 
\end{equation}

where $h^0, h^i, h^{ij}$ are the parameters and $i, j$ are the qubit indices. The expected values received after subsequent measurement allow to train the model. Local cost function calculates the output error of each specific operation based on these expected values. In the issue quantum and classical networks performed almost the same. 

Using the trigonometric property Schuld \cite{schuld2019evaluating}, Crooks \cite{crooks2019gradients}, Banchi \cite{banchi2021measuring} showed another gradient calculation approach named parameter-shift rule.
In short, the rule provides the parameter gradient calculation as follows:

\begin{equation}
    \nabla_\theta f(x; \theta) = \frac{1}{2} \Big[f(x; \theta + \frac{\pi}{2}) - f(x, \theta - \frac{\pi}{2})\Big].
\end{equation}

Hence, function $f(x;\theta)$ updates parameters $\theta$ for the current input data $x$ and current parameter $\theta_i$.

Overall, researchers progressed in classical CNN imitating. In the first step, amplitude input data representation performs alike classical data representation. In the same manner, convolution layers process the input data and emit a single value either for each wire or range of wires. Similarly to classical pooling, quantum pooling layers are aimed to shrink the number of wires. Layer-wise measurements allow to update parameters equivalently to the classical updating process. In this work I construct the training pipeline that leads mean squared error (MSE) to the lowest and most stable value.

\section{Problem statement}
\label{sec:problem}
\begin{figure}[!ht]
    \centering
        \begin{subfigure}[b]{0.25\textwidth}
            
        \end{subfigure}
        \hfill
        \begin{subfigure}[b]{0.25\textwidth}
            \centering
             \includegraphics[width=\textwidth]{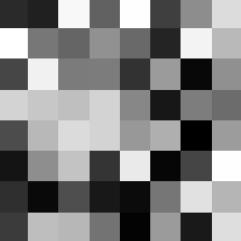}
             \caption{noised image}
             \label{fig:task_noised}
        \end{subfigure}
        \hfill
        \begin{subfigure}[b]{0.25\textwidth}
             \centering
             \includegraphics[width=\textwidth]{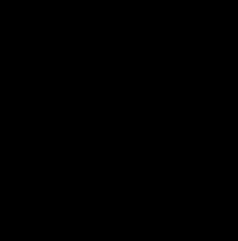}
             \caption{single colored image}
             \label{fig:task_pure}
         \end{subfigure}
         \hfill
         \begin{subfigure}[b]{0.25\textwidth}
            
        \end{subfigure}
    \caption{Training images:~\ref{fig:task_noised} is noised (with label 0) and~\ref{fig:task_pure} is a single-colored (with label 1)} 
    \label{fig:task_imgs}
\end{figure}

Classification in machine learning is an image-wise label prediction \cite{michie1994machine}. Binary classification limits label set to the values \{0, 1\}. Current outline aims to classify noised image (label $0$) and pure single-colored image (label $1$). For each image the pixel intensity varies in the set $\{0, 255\}$. Figure~\ref{fig:task_imgs} provides dataset samples from each class.

Image generator provides images randomly, thus, number of samples from both categories are almost equal.

\section{Solution}
\label{sec:solution}
Current work examines three QCNN architectures. The first task is to compare classical single-layered \textbf{conv} network results with quantum for the image of size $2\times2$. Then \textbf{conv-pool-pool} quantum model trains on images with $4\times4$ pixels. The final \textbf{conv-pool-conv-pool} quantum model processes an $8\times8$ image. Further sections explain quantum convolutional neural network structures in details.

\subsection{Quantum gates}
\label{sec:quantum_gates}
In the current work quantum circuit contains four different gates: $R_x(\theta), CX, CY, CZ$.
The $R_x(\theta)$ gate represents the rotation about X axis by angle $\alpha$ and described as follows:

\begin{equation}
    R_x(\theta) = 
    \begin{bmatrix}
        \cos{\frac{\theta}{2}} & -i\sin{\frac{\theta}{2}} \\
        -i\sin{\frac{\theta}{2}} & \cos{\frac{\theta}{2}}
    \end{bmatrix}.
\end{equation}

The family of $C_{x,y,z}$ gates represent the controlled flip operation about $x$, $y$, or $z$ axis correspondingly. Equations below explain these gates mathematically.

\begin{equation}
    CX = 
    \begin{bmatrix}
        1 & 0 & 0 & 0 \\
        0 & 1 & 0 & 0 \\
        0 & 0 & 1 & 0 \\
        0 & 0 & 0 & -1
    \end{bmatrix},
    \qquad
    CY =
    \begin{bmatrix}
        1 & 0 & 0 & 0 \\
        0 & 0 & 0 & -i \\
        0 & 0 & 1 & 0 \\
        0 & i & 0 & 0
    \end{bmatrix},
    \qquad
    CZ = 
    \begin{bmatrix}
        1 & 0 & 0 & 0 \\
        0 & 0 & 0 & 1 \\
        0 & 0 & 1 & 0 \\
        0 & 1 & 0 & 0
    \end{bmatrix}.
\label{eq:gates}
\end{equation}

Vector $\begin{bmatrix} \alpha & \beta \end{bmatrix}^T$ describes each qubit state under condition $|\alpha|^2 + |\beta|^2 = 1$. In the scope of this work, the maximum number of qubits utilized simultaneously equals two. The state of a qubit couple is $\begin{bmatrix} \alpha_1\cdot\alpha_2 & \alpha_1\cdot\beta_2 & \beta_1\cdot\alpha_2 & \beta_1\cdot\beta_2 \end{bmatrix}^T$.
Therefore, quantum circuit is a matrix multiplication processing the input vector. The input vector is a flatten lattice of the classical image pixels. Next section tells how to encode the image properly. 

\subsection{Data encoding}
\label{sec:data_ecoding}
\begin{figure}[!ht]
    \centering
            \begin{subfigure}{0.25\textwidth}
                
            \end{subfigure}
            \hfill
            \begin{subfigure}[b]{0.25\textwidth}
                 \begin{tikzpicture}
                    \node[anchor=center,inner sep=0] at (0,0) {\includegraphics[width=\textwidth]{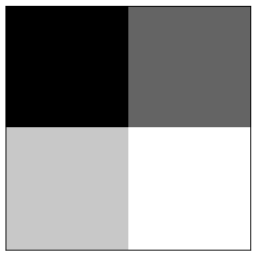}};
                    \node[matrix of math nodes, anchor=center, inner sep=0pt,
                        nodes={yshift = 0, minimum size=2cm, anchor=center}, color=white] at (0,0)
                        {0 & 125\\
                        \ & \\};
                    \node[matrix of math nodes, anchor=center, inner sep=0pt,
                        nodes={yshift = 0, minimum size=2cm, anchor=center}, color=black] at (0,0)
                        { & \\
                        200 & 255\\};
                \end{tikzpicture}
                \caption{Classical image}
                \label{fig:Cl_im}
            \end{subfigure}
            \hfill
            \begin{subfigure}[b]{0.25\textwidth}
                \begin{quantikz}
                    \lstick{$\ket{0}$} & \gate{R_{y}\frac{0\pi}{255}} & \meter{}\\
                    \lstick{$\ket{0}$} & \gate{R_{y}\frac{125\pi}{255}} & \meter{}\\
                    \lstick{$\ket{0}$} & \gate{R_{y}\frac{200\pi}{255}} & \meter{}\\
                    \lstick{$\ket{0}$} & \gate{R_{y}\frac{255\pi}{255}} & \meter{}\\
                \end{quantikz} 
                \caption{Qubit Lattice}
                \label{fig:Qubit_Lattice}
            \end{subfigure}
            \hfill
            \begin{subfigure}{0.25\textwidth}
                
            \end{subfigure}
    \caption{Qubit Lattice classical image encoding. Figure~\ref{fig:Cl_im} shows the classical $2\times2$ gray-scaled image. Figure~\ref{fig:Qubit_Lattice} represents the Qubit Lattice encoding. The representation circuit contains the same number of wires that image number of pixels. The rotation gates allow to set the state for each wire separately}
    \label{fig:qubit_lattice}
\end{figure}

Quantum neural network assumes quantum data encoding. Different image encoding algorithms exist \cite{latorre2005image, le2011flexible, sun2011multi, li2013image, yuan2014sqr, li2014multi, jiang2015quantum, csahin2018qrmw, li2018quantum, xu2019order, wang2020quantum, grigoryan2020new}. The Qubit Lattice \cite{venegas2003storing} is the simplest and the closest to the classical representation technique. The amplitude angles of range $\{0, \pi\}$ represent classical pixels intensity of range $\{0, 255\}$ correspondingly. The figure~\ref{fig:qubit_lattice} shows an image example and its encoding circuit.

After encoding, images passes through the QCNN. This step is aimed to extract features significant for classification. Convolution operation aims to discover these features.

\subsection{Convolution and pooling layers}
\label{sec:conv_and_pool}
\begin{figure}[!ht]
    \centering
        \begin{subfigure}[b]{0.25\textwidth}
            
        \end{subfigure}
        \hfill
        \begin{subfigure}[b]{0.25\textwidth}
            \centering
             \includegraphics[width=\textwidth]{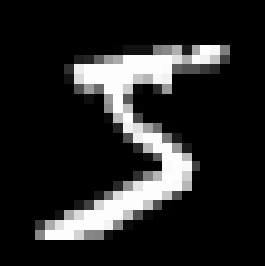}
             \caption{original sample}
             \label{fig:original_sample}
        \end{subfigure}
        \hfill
        \begin{subfigure}[b]{0.25\textwidth}
             \centering
             \includegraphics[width=\textwidth]{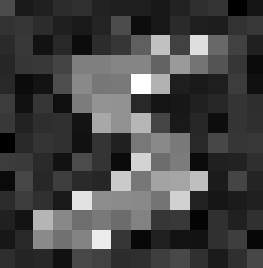}
             \caption{qconv result}
             \label{fig:qconv_result}
         \end{subfigure}
         \hfill
         \begin{subfigure}[b]{0.25\textwidth}
            
        \end{subfigure}
    \caption{MNIST original sample~\ref{fig:original_sample} and a result after convolution operation~\ref{fig:qconv_result}} 
    \label{fig:sample_conv}
\end{figure}

The convolution kernel from \cite{oh2020tutorial} bases convolution layers for neural networks is case of this work. As far as author does not provide angle information for $RX$ and $RZ$ gates I replaced $RX$ by $CY$, $RZ$ by $CZ$. Each kernel has size $2\times2$, stride two, and four trainable hyper-parameters. A single layer processing provides the result on a MNIST sample as in the figure~\ref{fig:sample_conv}.

\begin{figure}[!ht]
    \centering
        \scalebox{0.55}
        {
            \begin{quantikz}
                \lstick{$\ket{a}$} & \gate{RX(a_{00})} & \gate{CZ} & \gate{CY} & \gate{CZ} & \gate{CY} & \gate{CX} & \gate{CX} & \meter{}\\
                \lstick{$\ket{b}$} & \gate{RX(a_{01})} & \ctrl{-1} & \ctrl{-1} & \rstick{} \\
                \lstick{$\ket{c}$} & \gate{RX(a_{10})} & \gate{CZ} & \gate{CY} & \ctrl{-2} & \ctrl{-2} & \rstick{}\\
                \lstick{$\ket{d}$} & \gate{RX(a_{11})} & \ctrl{-1} & \ctrl{-1} & \rstick{} \\
                \lstick{$\ket{e}$} & \gate{RX(a_{00})} & \gate{CZ} & \gate{CY} & \gate{CZ} & \gate{CY} & \ctrl{-4} & \rstick{}\\
                \lstick{$\ket{f}$} & \gate{RX(a_{01})} & \ctrl{-1} & \ctrl{-1} & \rstick{} \\
                \lstick{$\ket{g}$} & \gate{RX(a_{10})} & \gate{CZ} & \gate{CY} & \ctrl{-2} & \ctrl{-2} & \rstick{}\\
                \lstick{$\ket{h}$} & \gate{RX(a_{11})} & \ctrl{-1} & \ctrl{-1} & \rstick{} \\
                \lstick{$\ket{i}$} & \gate{RX(a_{00})} & \gate{CZ} & \gate{CY} & \gate{CZ} & \gate{CY} & \gate{CX} & \ctrl{-8} & \rstick{}\\
                \lstick{$\ket{j}$} & \gate{RX(a_{01})} & \ctrl{-1} & \ctrl{-1} & \rstick{} \\
                \lstick{$\ket{k}$} & \gate{RX(a_{10})} & \gate{CZ} & \gate{CY} & \ctrl{-2} & \ctrl{-2} & \rstick{}\\
                \lstick{$\ket{l}$} & \gate{RX(a_{11})} & \ctrl{-1} & \ctrl{-1} & \rstick{} \\
                \lstick{$\ket{m}$} & \gate{RX(a_{00})} & \gate{CZ} & \gate{CY} & \gate{CZ} & \gate{CY} & \ctrl{-4} & \rstick{}\\
                \lstick{$\ket{n}$} & \gate{RX(a_{01})} & \ctrl{-1} & \ctrl{-1} & \rstick{}\\
                \lstick{$\ket{o}$} & \gate{RX(a_{10})} & \gate{CZ} & \gate{CY} & \ctrl{-2} & \ctrl{-2} & \rstick{}\\
                \lstick{$\ket{p}$} & \gate{RX(a_{11})} & \ctrl{-1} & \ctrl{-1} & \rstick{}
            \end{quantikz}
        }
    \caption{Quantum neural network for $4\times4$ image. Each $\ket{a}\ldots\ket{p}$ represents pixel intensity amplitude in the Qubit Lattice representation. Six rotation gates form a convolution-pooling layer. Each layer handles every four wires is the same manner. Four angles $a_{00}, a_{01}, a_{10}, a_{11}$ out of six are trainable}
    \label{fig:4_by_4_NN}
\end{figure}
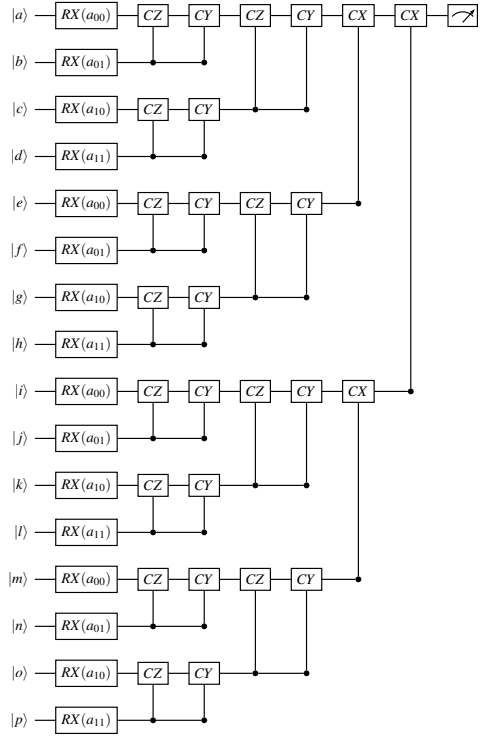

While convolution layers extract features, pooling shrinks the amount of data. Two different quantum pooling types exist: with and without error-correction. To simplify the task, I use pooling without error correction. Circuit~\ref{fig:4_by_4_NN} contains two pooling layers after a single convolution. 

Usually, convolution layers alternate with pooling layers to form a CNN. In this outline processing circuit for the image of $4\times4=16$ pixels contains \textbf{conv-pool-pool} structure; circuit for the image of $8\times8=64$ pixels contains \textbf{conv-pool-conv-pool} structure.

After passing whole network, image collapses into a single value - image label. In case of quantum computations, single value means single qubit measured multiple times. In order to train the network, loss function computes the error. Section below explains this process in particular.  

\subsection{Measurement and error computation}
\label{sec:measurement_and_error}

The term "measurement" means switching from quantum to classical data. Repeated circuit executions (shots) and measurements lead to a stable results. In this research number of shots equals 1000 for a single output wire to obtain a classification result.

In the current report back-propagation technique depends on the derivative calculation approach. Logistic activation function and circuit function derivative forms two back-propagation approaches. The third approach is a combination of the first two.

The logistic function is differentiable and widely used in classical implementations. The function equation is the following:

\begin{equation}
    \sigma(x) = \frac{1}{1+e^{-x}}.
\end{equation}

The circuit output measurement takes a single value from the set $\{0,1\}$. And the $arccos$ value from multiple measurements expresses the probability of getting label $0$ or $1$. On the one hand, the logistic function allows to estimate the final label prediction. On the other hand, the quantum circuit is derivative itself. Thus, guess to get rid of activation function non-linearity comes. Parameter-shift rule described in section~\ref{sec:rel_work} allows to check this assumption.

Next section describes three different back-propagation techniques depending on each derivative computation.

\subsection{Back-propagation}
\label{sec:back_propagation}
Introduction~\ref{sec:rel_work} mentioned that authors redraw the classical CNN with quantum mathematics. Measuring output after each layers allows to back-propagate the output error easily in the same classical manner \cite{zhang2016derivation}. This approach implies activation function derivative calculation.

At the same time, measurements destroy the quantum states to classical. Consequently, end-to-end data processing is more natural. This approach implies intermediate measurements reduction. In such case, only circuit derivative (shift-rule) is prominent. Further results show that it is empirically correct.

The final derivative calculation approach combines first two.

In the first assumption, logistic function derivative updates weight represented by gates angles. The equation below describes this derivative:

\begin{equation}
   \sigma'(x) = \sigma(x)\cdot(1-\sigma(x)).
\end{equation}

The following equation describes the updating process:

\begin{equation}
    params = params + l_{rate} \times(conv_{outs}^T \cdot error \times{pool_{sigmoidDerivative}})^T,
\end{equation}

where $params$ are the current angle values,  $l_{rate}$ is a learning rate, $conv_{outs}$ are the circuit outputs, $pool_{sigmoidDerivative}$ are the sigmoid derivatives for each sample in batch.

The second assumption says, repeated circuit executions and measurements allow to update parameters for each convolution separately. To do this, circuit executes $(n+1)\times N$, where $n$ is a total number on convolution layer, $1$~stands for circuit execution without any parameters change, $N$ is a number of shots.

Depending on number of layers amount of hyper-parameters and shots changes. Additionally, such parameter as learning rate also affects on the time consumption. Next sections provide the training parameters data and represent the result of all tested hypotheses.

\subsection{Neural networks parameters}
\label{sec:parameters}
Paper \cite{albawi2017understanding} explains the convolutional neural network terms including parameters. The following table~\ref{tab:nn_params} contains neural network parameters information for the current research. Observations allowed to estimate epoch number, learning rate and batch size empirically.

\begin{table}[!ht]
    \centering
    \begin{tabular}{c|c}
     \textbf{Parameter} & \textbf{Value} \\ [0.5ex] 
     \hline
     Convolution stride & 2  \\ 
     \hline
     Convolution kernel & $2\times2$ \\
     \hline
     Pooling stride & 2 \\
     \hline
     Pooling kernel & $1\times2$ \\
     \hline
     Batch size & 1000 \\
     \hline
     Learning rate & $10^{-7}$ \\
     \hline
     Epochs & 500 \\
     \hline
     Metric & Mean squared error (MSE) \\ [1ex] 
     \hline
    \end{tabular}
    \caption{Quantum neural network parameters}
    \label{tab:nn_params}
\end{table}

Each training process contains mentioned parameters. The next section describes the achieved results.

\section{Results}
\label{sec:results}
Convolution and pooling layers combinations form different neural network configurations.
The network structures within the current framework are following:
\begin{enumerate}
    \item conv;
    \item conv-pool-pool;
    \item conv-pool-conv-pool.
\end{enumerate}

\begin{figure}[!ht]
    \centering
    \begin{subfigure}[b]{0.45\textwidth}
        \scalebox{0.8}
        {
            \begin{tikzpicture}
                \begin{axis}[ylabel=MSE,
                             xlabel=Epoch,
                             ymin=0.21,
                             ymax=0.36,
                             title={CNN},
                             ]
                \addplot [black] table [x=x, y=classical, col sep=comma] {graphs/MSE.csv};
                \end{axis}
            \end{tikzpicture}
        }
        \caption{Classical NN}
        \label{fig:class_conv}
    \end{subfigure}
    \hfill
    \begin{subfigure}[b]{0.45\textwidth}
        \scalebox{0.8}
        {
            \begin{tikzpicture}
                \begin{axis}[ylabel=MSE,
                             xlabel=Epoch,
                             ymin=0.21,
                             ymax=0.36,
                             title={QCNN},
                             ]
                \addplot [black] table [x=x, y=conv_quantum_7, col sep=comma] {graphs/MSE.csv};
                \end{axis}
            \end{tikzpicture}
        }
        \caption{QCNN with circuit derivative}
        \label{fig:q_conv_circ}
    \end{subfigure}
    \begin{subfigure}[b]{0.45\textwidth}
        \scalebox{0.8}
        {
            \begin{tikzpicture}
                \begin{axis}[ylabel=MSE,
                             xlabel=Epoch,
                             ymin=0.21,
                             ymax=0.36,
                             title={QCNN},
                             ]
                \addplot [black] table [x=x, y=conv_sigma_7, col sep=comma] {graphs/MSE.csv};
                \end{axis}
            \end{tikzpicture}
        }
        \caption{QCNN with sigma derivative NN}
        \label{fig:q_conv_sigma}
    \end{subfigure}
    \hfill
    \begin{subfigure}[b]{0.45\textwidth}
        \scalebox{0.8}
        {
            \begin{tikzpicture}
                \begin{axis}[ylabel=MSE,
                             xlabel=Epoch,
                             ymin=0.21,
                             ymax=0.36,
                             title={QCNN},
                             ]
                \addplot [black] table [x=x, y=conv_both_7, col sep=comma] {graphs/MSE.csv};
                \end{axis}
            \end{tikzpicture}
        }
        \caption{QCNN with circuit and sigma derivatives}
        \label{fig:q_conv_both}
    \end{subfigure}
    \caption{MSE function results depending on the gradient calculation algorithm, learning rate equals to $10^{-7}$}
    \label{fig:conv_results}
\end{figure}
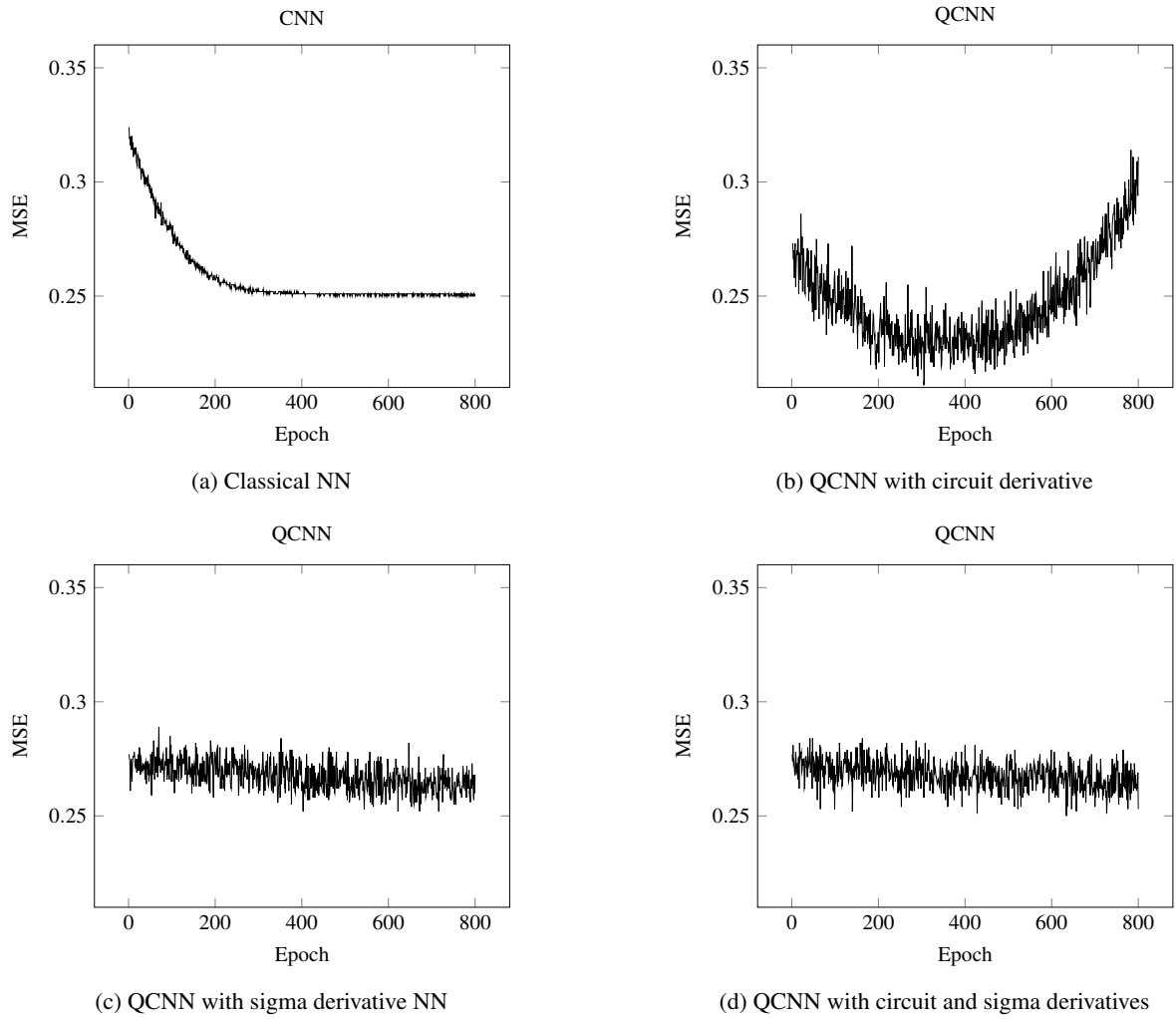

The first observation compares classical and quantum single-layered network in order to confirm the obtained results. Figure~\ref{fig:conv_results} compares error of classical and quantum convolutional neural networks.

\begin{figure}[!ht]
    \centering
    \begin{subfigure}[b]{0.45\textwidth}
        \scalebox{0.8}
        {
            \begin{tikzpicture}
                \begin{axis}[ylabel=MSE,
                             xlabel=Epoch,
                             ymin=0.21,
                             ymax=0.36,
                             title={QCNN},
                             ]
                \addplot [black] table [x=x, y=conv_sigma_6, col sep=comma] {graphs/MSE.csv};
                \end{axis}
            \end{tikzpicture}
        }
        \caption{$l_{rate} = 10^{-6}$}
        \label{fig:q_conv_both_6}
    \end{subfigure}
    \hfill
    \begin{subfigure}[b]{0.45\textwidth}
    \scalebox{0.8}
        {
            \begin{tikzpicture}
                \begin{axis}[ylabel=MSE,
                             xlabel=Epoch,
                             ymin=0.21,
                             ymax=0.36,
                             title={QCNN},
                             ]
                \addplot [black] table [x=x, y=conv_sigma_8, col sep=comma] {graphs/MSE.csv};
                \end{axis}
            \end{tikzpicture}
        }
        \caption{$l_{rate} = 10^{-8}$}
        \label{fig:q_conv_both_8}
    \end{subfigure}
    \caption{MSE function results depending on the learning rate. While figure~\ref{fig:q_conv_both_6} represents the unstable training behaviour, figure~\ref{fig:q_conv_both_8} shows the loss function settling.}
    \label{fig:q_conv_lr}
\end{figure}
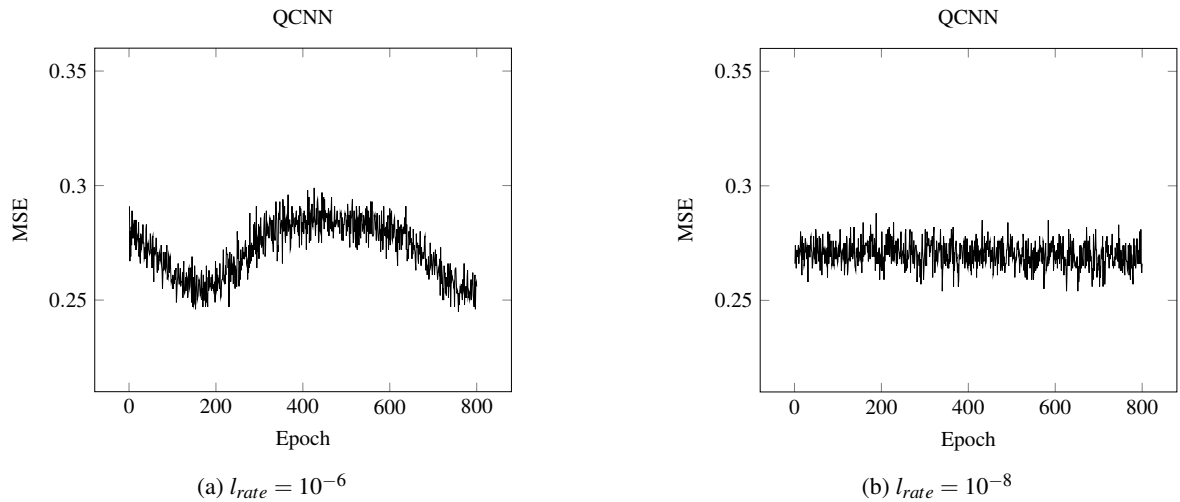

The model is sensitive to the learning rate upper bound. Figure~\ref{fig:q_conv_lr} shows the results with learning rates $10^{-6}$ and $10^{-8}$.

Such sensitive behaviour is common for all the three cases within current experiment. To reduce the learning rate means to set gradient incremental steps smaller. The tinier gradient value, the more reliable but more time-consuming the learning is. At the same time, difference between MSE indicators in the figure~\ref{fig:q_conv_both} and ~\ref{fig:q_conv_both_8} visually is not significant. Moreover, each QCNN training process consumes 4~hours~37~minutes. Due to high time consumption, learning rate with value $10^{-7}$ is supposed to be optimal.

From the visual point of view, the training with circuit derivative is closer to the classical training process. The loss function is gradually settling from the maximum values on the first iterations to the minima. Additionally, this minima ($0.231$) is lower than in the classical case ($0.255$). Thus, training process based on the parameter-shift rule shows the better result within all three approaches.

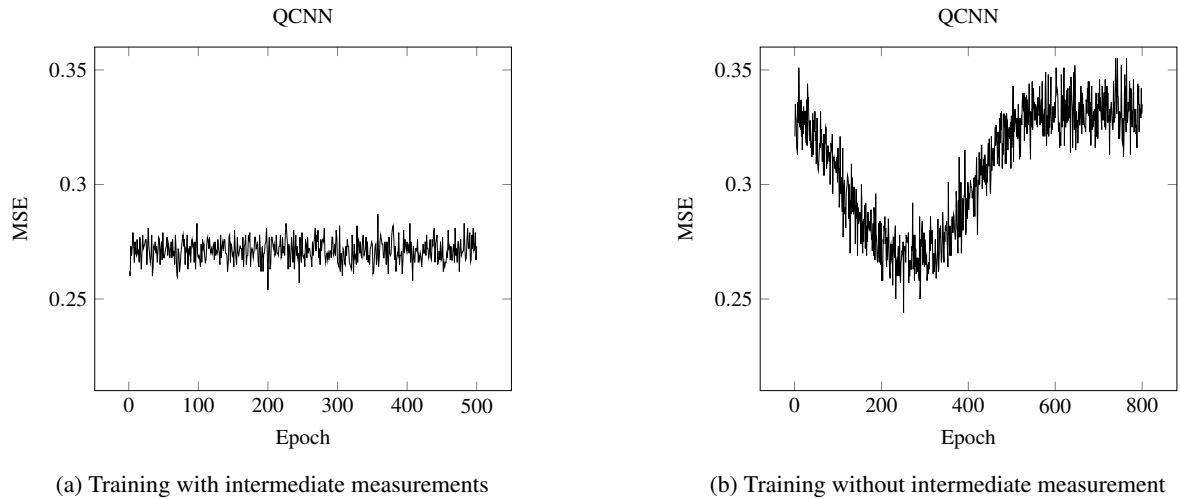
\begin{figure}[!ht]
    \centering
    \begin{subfigure}[b]{0.45\textwidth}
        \scalebox{0.8}
        {
            \begin{tikzpicture}
                \begin{axis}[ylabel=MSE,
                             xlabel=Epoch,
                             ymin=0.21,
                             ymax=0.36,
                             title={QCNN}
                             ]
                \addplot [black] table [x=x, y=pool_measuered, col sep=comma] {graphs/MSE.csv};
                \end{axis}
            \end{tikzpicture}
        }
        \caption{Training with intermediate measurements}
        \label{fig:q_conv_pool_pool_no_measure}
    \end{subfigure}
    \hfill
    \begin{subfigure}[b]{0.45\textwidth}
        \scalebox{0.8}
        {
            \begin{tikzpicture}
                \begin{axis}[ylabel=MSE,
                             xlabel=Epoch,
                             ymin=0.21,
                             ymax=0.36,
                             title={QCNN}
                             ]
                \addplot [black] table [x=x, y=pool_without, col sep=comma] {graphs/MSE.csv};
                \end{axis}
            \end{tikzpicture}
        }
        \caption{Training without intermediate measurement}
        \label{fig:q_conv_pool_pool_with_measure}
    \end{subfigure}
    \caption{MSE function results depending on the measurements in the training process. Figure \ref{fig:q_conv_pool_pool_no_measure} represents results of layer-wise measured circuit and Figure \ref{fig:q_conv_pool_pool_with_measure} shows the end-to-end circuit with only measurement in the end}
    \label{fig:q_conv_pool_pool_training}
\end{figure}

The next experiment towards \textbf{conv-pool-pool} neural network aims to check whether intermediate measurements affect on the results. Measuring convolution output wires switches the quantum state to the classical from the set $\{0, 1\}$. Iterative convolution circuit execution allows to get statistics of how much each wire is close to $0$ and $1$. Statistical value from each value represents an input value to the pooling layer. Then pooling output is measured again. Overall, the experiment leads to the results as in the figure~\ref{fig:q_conv_pool_pool_training}.

The circuit sensitivity to the learning rate appears also here. Despite the learning rate stays remain $10^{-7}$, the training process did not stabilize. However, metric values roll from maxima to the minimum value while model without measurements is training. Oppositely,  model with measurement behaves almost statically. About time, model without measurement training takes 18 hours 39 minutes, the model with measurement training takes 18 hours 32 minutes.

In this experiment model without intermediate measurements shows classically-looking training behaviour. Thus, I assume that end-to-end circuit is preferable. The last experiment is based on this assumption.

\begin{figure}[!ht]
    \centering
    \begin{subfigure}[b]{0.45\textwidth}
    \scalebox{0.8}
            {
                \begin{tikzpicture}
                    \begin{axis}[ylabel=MSE,
                                 xlabel=Epoch,
                                 ymin=0.21,
                                 ymax=0.36,
                                 title={QCNN}
                                 ]
                    \addplot [black] table [x=x, y=multi_layer_wise, col sep=comma] {graphs/MSE.csv};
                    \end{axis}
                \end{tikzpicture}
            }
        \caption{Training with layer-wise weights update}
        \label{fig:q_conv_pool_conv_pool_layerwise}
    \end{subfigure}
    \hfill
    \begin{subfigure}[b]{0.45\textwidth}
        \scalebox{0.8}
        {
            \begin{tikzpicture}
                \begin{axis}[ylabel=MSE,
                             xlabel=Epoch,
                             ymin=0.21,
                             ymax=0.36,
                             title={QCNN}
                             ]
                \addplot [black] table [x=x, y=multi_no_measure, col sep=comma] {graphs/MSE.csv};
                \end{axis}
            \end{tikzpicture}
        }
        \caption{Training with simultaneously weights update}
        \label{fig:q_conv_pool_conv_pool_simultaneously}
    \end{subfigure}
    \caption{MSE function results depending on the weights update strategy. Figure \ref{fig:q_conv_pool_conv_pool_layerwise} represents results of layer-wise weights update and Figure \ref{fig:q_conv_pool_conv_pool_simultaneously} stands for training with simultaneous weights update}
    \label{fig:q_conv_pool_conv_pool_training}
\end{figure}
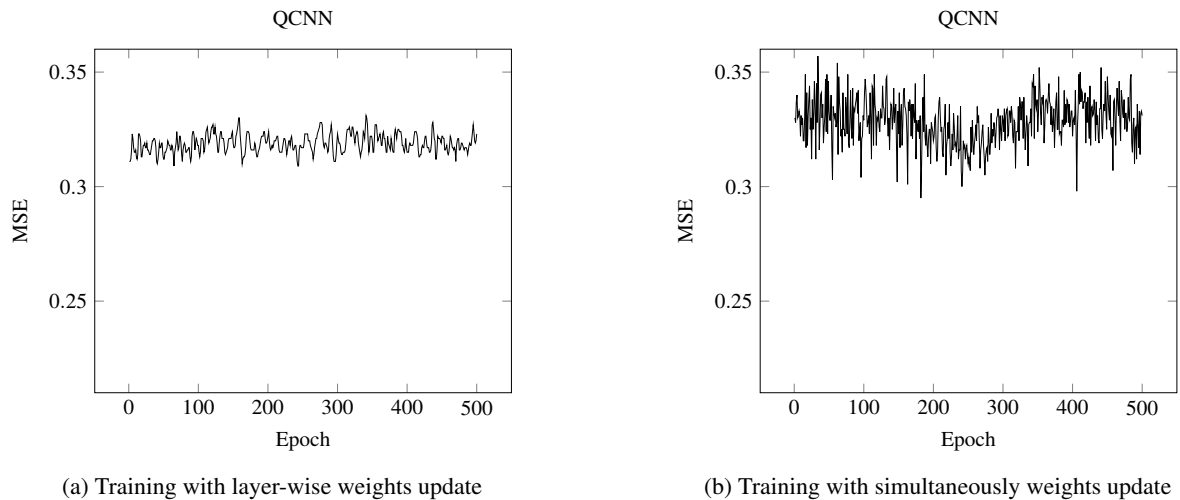

The last model has \textbf{conv-pool-conv-pool} configuration. The training purpose is to check either layer-wise or simultaneous weights update is preferable. Figure~\ref{fig:q_conv_pool_conv_pool_training} represents the experiments metric results.

In spite logical assumptions, the simultaneous weights update leads to more natural graph of the loss function. The function did not converge, however minima is visually observable. Inappropriate learning rate, large or oppositely not enough batch number or simply unsuccessful experiment - these are all possible reasons of the non-convergence. Unfortunately, the time that both models consume on training (115 hours 32 minutes for simultaneous weights update and 160 hours 37 minutes for layer-wise update models) make further experiments difficult.

Overall, during the discussed experiments I observed that quantum convolutional neural network training is successful when calculating derivative from the \textbf{end-to-end without intermediate measurements circuit via simultaneous weights update}. It is hard to clarify details as far as experiments are timely expensive.

\section{Discussion and conclusion}
\label{sec:discuss_and_conclude}
Current work describes convolutions neural networks training experience for binary image classification. To build, train, and evaluate end-to-end quantum neural networks is possible. The worst thing for developing those ideas is a high time-consumption. For example, to establish appropriate learning rate, multiple training takes weeks very easy. The possible solution could be batch number decreasing (thus, poor result quality), squeezing number of qubits (thus, new convolution and pooling blocks releasing) or wire number reduction (what causes above number of qubits problem).

I trained the networks with the parameters named in table \ref{tab:nn_params} and present the repository with the code. As a result, the network performs with the minimum MSE $0.23$ in the best case and 0.25 in case of classical model. I showed that large training time dramatically influences on investigating quantum neural networks in my case.

This major problem may be addressed to the parallel data encoding (NEQR \cite{zhang2013neqr} as an example), training process optimization, new weights updating techniques.

\bibliographystyle{ieeetr}
\bibliography{OSAmeetings}

\end{document}